\documentstyle[12pt,axodraw]{article}

\catcode`@=12
\begin{document}
\thispagestyle{empty}
\begin{flushright} 
UCRHEP-T308\\ 
June 2001\
\end{flushright}
\vspace{0.5in}
\begin{center}
{\LARGE	\bf Softly Broken A$_4$ Symmetry for\\ Nearly Degenerate Neutrino 
Masses\\}
\vspace{1.5in}
{\bf Ernest Ma$^a$ and G. Rajasekaran$^b$\\}
\vspace{0.2in}
{\sl $^a$ Physics Department, University of California, Riverside, 
California 92521\\}
\vspace{0.1in}
{\sl $^b$ Institute of Mathematical Sciences, Chennai (Madras) 600113, India\\}
\vspace{1.5in}
\end{center}
\begin{abstract}\
The leptonic Higgs doublet model of neutrino masses is implemented with an 
$A_4$ discrete symmetry (the even permutation of 4 objects or equivalently 
the symmetry of the tetrahedron) which has 4 irreducible representations: 
\underline {1}, \underline {1}$'$, \underline {1}$''$, and \underline {3}.  
The resulting spontaneous and soft breaking of $A_4$ provides a realistic 
model of charged-lepton masses as well as a nearly degenerate neutrino mass 
matrix.  Phenomenological consequences at and below the TeV scale are 
discussed.
\end{abstract}
\newpage
\baselineskip 24pt

\section{Introduction}
Since the experimental evidence of neutrino oscillations \cite{atm,solar,lsnd} 
requires only neutrino mass differences, the possibility of nearly degenerate 
neutrino masses is often considered \cite{amr}.  However, the charged lepton 
masses are certainly not degenerate, so whatever symmetry we use to maintain 
the neutrino mass degeneracy must be broken.  To implement this idea in a 
renormalizable field theory, the symmetry in question should be broken only 
spontaneously and by explicit soft terms (if it is not a gauge symmetry).

Recently, a simple model of neutrino masses was proposed \cite{ma01} using 
a leptonic Higgs doublet $\eta = (\eta^+,\eta^0)$ and 3 right-handed singlet 
fermions $N_{iR}$, all of which are at or below the TeV energy scale. 
It was further shown \cite{mr} that this model is able to account for the 
recent measurement \cite{g-2} of the muon anomalous magnetic moment, provided 
that neutrino masses are nearly degenerate \cite{tmg}.

In this paper, the specific choice of a discrete symmetry, i.e. $A_4$ which 
is the symmetry group of the even permutation of 4 objects or equivalently 
that of the tetrahedron, is used to 
sustain this degeneracy, which is then broken both spontaneously to 
generate the different charged-lepton masses, and softly to account for the 
mass splitting and mixing of the neutrinos.  In Section 2, the group $A_4$ 
and its irreducible representations are discussed.  In Section 3, the 
structure of the leptonic model, which has altogether 4 Higgs doublets, 
is presented.  In Section 4, phenomenological consequences of this model 
are explored.  In Section 5, the quark sector is discussed. 
In Section 6, there are some concluding remarks.

\section{Discrete Symmetry A$_4$}

The finite group of the even permutation of 4 objects, i.e. $A_4$, has 12 
elements, which are divided into 4 classes, with number of elements 1,4,4,3 
respectively.  This means that there are 4 irreducible representations, with 
dimensions $n_i$, such that $\sum_i n_i^2 = 12$.  There is only one solution: 
$n_1 = n_2 = n_3 = 1$ and $n_4 = 3$, and the character table of the 4 
representations is given below.

\begin{table}[htb]
\caption{Character Table of $A_4$}
\begin{center}
\begin{tabular}{|c|c|c|c|c|}
\hline
Class & $\chi^{(1)}$ & $\chi^{(2)}$ & $\chi^{(3)}$ & $\chi^{(4)}$ \\ 
\hline 
$C_1$ & 1 & 1 & 1 & 3 \\ 
$C_2$ & 1 & $\omega$ & $\omega^2$ & 0 \\ 
$C_3$ & 1 & $\omega^2$ & $\omega$ & 0 \\ 
$C_4$ & 1 & 1 & 1 & --1 \\
\hline 
\end{tabular}
\end{center}
\end{table}

The complex number $\omega$ is the cube root of unity, i.e. $e^{2 \pi i/3}$. 
Hence $1 + \omega + \omega^2 = 0$.  Calling the 4 irreducible representations 
\underline {1}, \underline {1}$'$, \underline {1}$''$, and \underline {3} 
respectively, we have the decomposition
\begin{equation}
\underline {3} \times \underline {3} = \underline {1} + \underline {1}' + 
\underline {1}'' + \underline {3} + \underline {3}.
\end{equation}
In particular, denoting \underline {3} as $(a,b,c)$, we have
\begin{eqnarray}
\underline {1} &=& a_1 a_2 + b_1 b_2 + c_1 c_2, \\ 
\underline {1}' &=& a_1 a_2 + \omega^2 b_1 b_2 + \omega c_1 c_2, \\ 
\underline {1}'' &=& a_1 a_2 + \omega b_1 b_2 + \omega^2 c_1 c_2.
\end{eqnarray}
For completeness, the $3 \times 3$ representation matrices of the 12 group 
elements are given below.
\begin{eqnarray}
C_1 &:& \left[ \begin{array} {c@{\quad}c@{\quad}c} 1 & 0 & 0 \\ 0 & 1 & 0 \\ 
0 & 0 & 1 \end{array} \right], \\ C_2 &:& \left[ \begin{array} 
{c@{\quad}c@{\quad}c} 0 & 0 & 1 \\ 1 & 0 & 0 \\ 0 & 1 & 0 \end{array} \right], 
\left[ \begin{array} {c@{\quad}c@{\quad}c} 0 & 0 & 1 \\ -1 & 0 & 0 \\ 
0 & -1 & 0 \end{array} \right], \left[ \begin{array} {c@{\quad}c@{\quad}c} 
0 & 0 & -1 \\ 1 & 0 & 0 \\ 0 & -1 & 0 \end{array} \right], \left[ 
\begin{array} {c@{\quad}c@{\quad}c} 0 & 0 & -1 \\ -1 & 0 & 0 \\ 0 & 1 & 0 
\end{array} \right], \\ C_3 &:& \left[ \begin{array} {c@{\quad}c@{\quad}c} 
0 & 1 & 0 \\ 0 & 0 & 1 \\ 1 & 0 & 0 \end{array} \right], \left[ \begin{array} 
{c@{\quad}c@{\quad}c} 0 & 1 & 0 \\ 0 & 0 & -1 \\ -1 & 0 & 0 \end{array} 
\right], \left[ \begin{array} {c@{\quad}c@{\quad}c} 0 & -1 & 0 \\ 0 & 0 & 1 
\\ -1 & 0 & 0 \end{array} \right], \left[ \begin{array} {c@{\quad}c@{\quad}c} 
0 & -1 & 0 \\ 0 & 0 & -1 \\ 1 & 0 & 0 \end{array} \right], \\ C_4 &:& 
\left[ \begin{array} {c@{\quad}c@{\quad}c} 1 & 0 & 0 \\ 0 & -1 & 0 \\ 
0 & 0 & -1 \end{array} \right], \left[ \begin{array} {c@{\quad}c@{\quad}c} 
-1 & 0 & 0 \\ 0 & 1 & 0 \\ 0 & 0 & -1 \end{array} \right], \left[ 
\begin{array} {c@{\quad}c@{\quad}c} -1 & 0 & 0 \\ 0 & -1 & 0 \\ 0 & 0 & 1 
\end{array} \right].
\end{eqnarray}

\section{Model of Nearly Degenerate Neutrino Masses}

Under $A_4$ and $L$ (lepton number), the color-singlet fermions and scalars 
of this model transform as follows.
\begin{eqnarray}
(\nu_i, l_i)_L &\sim& (\underline {3}, 1), \\ 
l_{1R} &\sim& (\underline {1}, 1), \\ 
l_{2R} &\sim& (\underline {1}', 1), \\ 
l_{3R} &\sim& (\underline {1}'', 1), \\
N_{iR} &\sim& (\underline {3}, 0), \\ 
\Phi_i = (\phi_i^+, \phi_i^0) &\sim& (\underline {3}, 0), \\ 
\eta = (\eta^+, \eta^0) &\sim& (\underline {1}, -1).
\end{eqnarray}
Hence its Lagrangian has the invariant terms
\begin{equation}
{1 \over 2} M N_{iR}^2 + f \bar N_{iR} (\nu_{iL} \eta^0 - l_{iL} \eta^+) + 
h_{ijk} \overline {(\nu_i,l_i)}_L l_{jR} \Phi_k + h.c.,
\end{equation}
where
\begin{equation}
h_{i1k} = h_1 \left[ \begin{array} {c@{\quad}c@{\quad}c} 1 & 0 & 0 \\ 
0 & 1 & 0 \\ 0 & 0 & 1 \end{array} \right], ~~ h_{i2k} = h_2 \left[ 
\begin{array} {c@{\quad}c@{\quad}c} 1 & 0 & 0 \\ 0 & \omega & 0 \\ 
0 & 0 & \omega^2 \end{array} \right], ~~ h_{i3k} = h_3 \left[ \begin{array} 
{c@{\quad}c@{\quad}c} 1 & 0 & 0 \\ 0 & \omega^2 & 0 \\ 0 & 0 & \omega 
\end{array} \right].
\end{equation}
Thus the neutrino mass matrix (in this basis) is proportional to the unit 
matrix with magnitude $f^2 u^2/M$, where $u = \langle \eta^0 \rangle$, 
whereas the charged-lepton mass matrix is given by
\begin{equation}
{\cal M}_l = \left[ \begin{array} {c@{\quad}c@{\quad}c} h_1 v_1 & h_2 v_1 & 
h_3 v_1 \\ h_1 v_2 & h_2 \omega v_2 & h_3 \omega^2 v_2 \\ h_1 v_3 & h_2 
\omega^2 v_3 & h_3 \omega v_3 \end{array} \right].
\end{equation}
If $v_1 = v_2 = v_3 = v$, then ${\cal M}_l$ is easily diagonalized:
\begin{equation}
U_L^\dagger {\cal M}_l U_R = \left[ \begin{array} {c@{\quad}c@{\quad}c} 
\sqrt 3 h_1 v & 0 & 0 \\ 0 & \sqrt 3 h_2 v & 0 \\ 0 & 0 & \sqrt 3 h_3 v 
\end{array} \right] = \left[ \begin{array} {c@{\quad}c@{\quad}c} m_e & 0 & 0 
\\ 0 & m_\mu & 0 \\ 0 & 0 & m_\tau \end{array} \right],
\end{equation}
where
\begin{equation}
U_L = {1 \over \sqrt 3} \left[ \begin{array} {c@{\quad}c@{\quad}c} 1 & 1 & 
1 \\ 1 & \omega & \omega^2 \\ 1 & \omega^2 & \omega \end{array} \right], ~~ 
U_R = \left[ \begin{array} {c@{\quad}c@{\quad}c} 1 & 0 & 0 \\ 0 & 1 & 0 \\ 
0 & 0 & 1 \end{array} \right].
\end{equation}
The $6 \times 6$ Majorana mass matrix spanning $(\bar \nu_e, \bar \nu_\mu, 
\bar \nu_\tau, N_1, N_2, N_3)$ is then given by
\begin{equation}
{\cal M}_{(\nu,N)} = \left[ \begin{array} {c@{\quad}c} 0 & U_L^\dagger f u \\ 
U_L^* f u & M \end{array} \right].
\end{equation}
Hence the $3 \times 3$ seesaw mass matrix for $(\nu_e, \nu_\mu, \nu_\tau)$ 
becomes
\begin{equation}
{\cal M}_\nu = {f^2 u^2 \over M} U_L^T U_L = {f^2 u^2 \over M} \left[ 
\begin{array} {c@{\quad}c@{\quad}c} 1 & 0 & 0 \\ 0 & 0 & 1 \\ 
0 & 1 & 0 \end{array} \right].
\end{equation}
This shows that $\nu_\mu$ mixes maximally with $\nu_\tau$, but since all 
physical neutrino masses are degenerate, there are no neutrino oscillations. 
To break the degeneracy, arbitrary soft terms of the form $m_{ij} N_{iR} 
N_{jR}$ may be added to Eq.~(16).  As a result, it is possible to have, 
for example, a bimaximal mixing pattern with the appropriate small neutrino 
mass-squared differences for atmospheric \cite{atm} and solar \cite{solar} 
neutrino oscillations used in Ref.[6].

\section{Phenomenological Consequences}

Whereas the minimal standard model has only one Higgs scalar doublet, our 
$A_4$ model has four, $\Phi_{1,2,3}$ and $\eta$.  The interplay between 
$\Phi_i$ and $\eta$ is the same as in Ref.[5], which allows $u = \langle 
\eta^0 \rangle$ to be small.  The new feature here is the structure 
of the Higgs sector containing $\Phi_i$.  The corresponding $A_4-$invariant 
Higgs potential is given by
\begin{eqnarray}
V &=& m^2 \sum_i \Phi_i^\dagger \Phi_i + {1 \over 2} \lambda_1 \left( 
\sum_i \Phi_i^\dagger \Phi_i \right)^2 \nonumber \\ &+& \lambda_2 
(\Phi_1^\dagger \Phi_1 + \omega^2 \Phi_2^\dagger \Phi_2 + \omega 
\Phi_3^\dagger \Phi_3)(\Phi_1^\dagger \Phi_1 + \omega \Phi_2^\dagger \Phi_2 
+ \omega^2 \Phi_3^\dagger \Phi_3) \nonumber \\ &+& \lambda_3 
[(\Phi_2^\dagger \Phi_3)(\Phi_3^\dagger \Phi_2) + (\Phi_3^\dagger \Phi_1) 
(\Phi_1^\dagger \Phi_3) + (\Phi_1^\dagger \Phi_2)(\Phi_2^\dagger \Phi_1)] 
\nonumber \\ &+& \left\{ {1 \over 2} \lambda_4 [(\Phi_2^\dagger \Phi_3)^2 + 
(\Phi_3^\dagger \Phi_1)^2 + (\Phi_1^\dagger \Phi_2)^2] + h.c. \right\}
\end{eqnarray}
Let $\langle \phi_i^0 \rangle = v_i$, then the minimum of $V$ is
\begin{eqnarray}
V_{min} &=& m^2 \left( |v_1|^2 + |v_2|^2 + |v_3|^2 \right) + {1 \over 2} 
\lambda_1 \left( |v_1|^2 + |v_2|^2 + |v_3|^2 \right)^2 \nonumber \\ 
&+& \lambda_2 \left( |v_1|^2 + \omega^2 |v_2|^2 + \omega |v_3|^2 \right) 
\left( |v_1|^2 + \omega |v_2|^2 + \omega^2 |v_3|^2 \right) \nonumber \\ 
&+& \lambda_3 \left( |v_2|^2 |v_3|^2 + |v_3|^2 |v_1|^2 + |v_1|^2 |v_2|^2 
\right) \nonumber \\ &+& \left\{ {1 \over 2} \lambda_4 \left[ (v_2^*)^2 v_3^2 
+ (v_3^*)^2 v_1^2 + (v_1^*)^2 v_2^2 \right] + c.c. \right\}
\end{eqnarray}
The minimization conditions on $v_i$ are given by
\begin{eqnarray}
0 = {\partial V_{min} \over \partial v_1^*} &=& m^2 v_1 + \lambda_1 v_1 
\left( |v_1|^2 + |v_2|^2 + |v_3|^2 \right) + \lambda_2 v_1 \left( 2 |v_1|^2 
- |v_2|^2 - |v_3|^2 \right) \nonumber \\ &+& \lambda_3 v_1 \left( |v_2|^2 
+ |v_3|^2 \right) + \lambda_4 v_1^* ( v_2^2 + v_3^2),
\end{eqnarray}
and other similar equations. Hence the solution
\begin{equation}
v_1 = v_2 = v_3 = v = \sqrt {-m^2 \over 3 \lambda_1 + 2 \lambda_3 + 2 
\lambda_4}
\end{equation}
is allowed if $\lambda_4$ is real.

The mass-squared matrices in the Re$\phi_i^0$, Im$\phi_i^0$, and 
$\phi_i^\pm$ bases are all of the form
\begin{equation}
{\cal M}^2 = \left( \begin{array} {c@{\quad}c@{\quad}c} a & b & b \\ 
b & a & b \\ b & b & a \end{array} \right),
\end{equation}
where
\begin{eqnarray}
{\rm Re}\phi_i^0: && a = 2(\lambda_1 + 2\lambda_2) v^2, ~~ b = 2(\lambda_1 
- \lambda_2 + \lambda_3 + \lambda_4) v^2, \\ 
{\rm Im}\phi_i^0: && a = -4 \lambda_4 v^2, ~~ b= 2 \lambda_4 v^2, \\ 
\phi_i^\pm: && a = -2 (\lambda_3 + \lambda_4) v^2, ~~ b = (\lambda_3 + 
\lambda_4) v^2.
\end{eqnarray}
The eigenvalues of ${\cal M}^2$ are $a + 2b$, $a-b$, and $a-b$.  Hence 
$(\Phi_1 + \Phi_2 + \Phi_3)/\sqrt 3$ has the properties of the standard-model 
Higgs doublet with mass-squared eigenvalues $2(3\lambda_1 + 2\lambda_3 + 
2\lambda_4) v^2$, 0, and 0 for Re$(\phi_1^0 + \phi_2^0 + 
\phi_3^0)/\sqrt 3$, Im$(\phi_1^0 + \phi_2^0 + \phi_3^0)/\sqrt 3$, and 
$(\phi_1^\pm + \phi_2^\pm + \phi_3^\pm)/\sqrt 3$ respectively. 
The two other linear combinations are mass-degenerate in each sector with 
mass-squared eigenvalues given by $M_R^2 = 2(3\lambda_2 - \lambda_3 - 
\lambda_4) v^2$, $M_I^2 = -6\lambda_4 v^2$, and $M_\pm^2 = -3(\lambda_3 + 
\lambda_4) v^2$ respectively.

The distinct phenomenological signatures of our $A_4$ model are thus given 
by the two new Higgs doublets.  They are predicted to be pairwise degenerate 
in mass and their Yukawa interactions are given by
\begin{eqnarray}
{\cal L}_{int} &=& \left( {m_\tau \over v} \overline {(\nu_e,e)}_L \tau_R + 
{m_\mu \over v} \overline {(\nu_\tau,\tau)}_L \mu_R + {m_e \over v} 
\overline {(\nu_\mu,\mu)}_L e_R \right) \Phi' \nonumber \\ 
&+& \left( {m_\tau \over v} \overline {(\nu_\mu,\mu)}_L \tau_R + 
{m_\mu \over v} \overline {(\nu_e,e)}_L \mu_R + {m_e \over v} \overline 
{(\nu_\tau,\tau)}_L e_R \right) \Phi'' + h.c.,
\end{eqnarray}
where
\begin{equation}
\Phi' = {1 \over \sqrt 3} (\Phi_1 + \omega \Phi_2 + \omega^2 \Phi_3), ~~~ 
\Phi'' = {1 \over \sqrt 3} (\Phi_1 + \omega^2 \Phi_2 + \omega \Phi_3).
\end{equation}
This means that lepton flavor is necessarily violated and serves as an 
unmistakable prediction of this model.

Using Eq.~(31), we find that the most prominent (with strength $m_\tau 
m_\mu/v^2$) exotic decays of this model are
\begin{equation}
\tau_R^- \to \mu_L^- \mu_R^- e_R^+,  ~~~ \tau_R^- \to \mu_L^- \mu_L^+ e_L^-,
\end{equation}
through $(\phi'')^0$ exchange.  The former amplitude is proportional to 
$M_0^{-2} = M_R^{-2} + M_I^{-2}$ and the latter to $M_1^{-2} = |M_R^{-2} - 
M_I^{-2}|$.  Hence
\begin{equation}
B(\tau^- \to \mu^- \mu^- e^+) = \left( {9 m_\tau^2 m_\mu^2 \over M_0^4} 
\right) \left( {v_0^2 \over 3 v^2} \right)^2 B(\tau \to \mu \nu \nu),
\end{equation}
where $v_0 = (2 \sqrt 2 G_F)^{-1/2}$ and $3 v^2 < v_0^2$.  Using 
$B(\tau \to \mu \nu \nu) = 0.174$, we find
\begin{equation}
B(\tau^- \to \mu^- \mu^- e^+) = 5.5 \times 10^{-10} \left( {v_0^2 \over 
3 v^2} \right)^2 \left( {100~{\rm GeV} \over M_0} \right)^4,
\end{equation}
as compared to the experimental upper bound of $1.5 \times 10^{-6}$. 
Similarly, $B(\tau^- \to \mu^- \mu^+ e^-)$ is also given by Eq.~(35) 
with $M_0$ replaced by $M_1$ (which is always greater than $M_0$), as 
compared to the experimental upper bound of $1.8 \times 10^{-6}$. 
Other $\tau$ decays are further suppressed because they are proportional 
to $m_\tau m_e$ or $m_\mu m_e$.  Note the important fact that there is 
no tree-level $\mu \to e e e$ decay in this model.

From Eq.~(31), there are also tree-level contributions to $\tau$ and $\mu$ 
decays through charged-scalar exchange.  For example,
\begin{equation}
\mu_R^- \to e_R^- \nu_\tau \bar \nu_\mu, ~~~ \mu_R^- \to e_R^- \nu_e 
\bar \nu_\tau,
\end{equation}
through $(\phi')^\pm$ and $(\phi'')^\pm$ exchange respectively.  However, 
these amplitudes are proportional to $m_\mu m_e$ and only add incoherently to 
the dominant $\mu_L^- \to e_L^- \nu_\mu \bar \nu_e$ amplitude.  Hence they 
are completely negligible.  The same holds true for $\tau$ decays, but to a 
lesser extent.

Consider next the muon anomalous magnetic moment, which receives a 
contribution proportional to $m_\tau^2$ from $(\phi'')^0$.  A straightforward 
calculation yields
\begin{equation}
\Delta a_\mu = {G_F m_\tau^2 \over 2 \sqrt 2 \pi^2} \left( {m_\mu^2 \over 
M_0^2} \right) \left( {v_0^2 \over 3 v^2} \right) = 1.5 \times 10^{-12} 
\left( {v_0^2 \over 3 v^2} \right) \left( {100~{\rm GeV} \over M_0} \right)^2,
\end{equation}
as compared to the possible discrepancy \cite{cm} of $(426 \pm 165) \times 
10^{-11}$, based on the recent experimental measurement \cite{g-2}.  Hence 
the contribution to $\Delta a_\mu$ from Eq.~(31) is negligible, and the 
latter's theoretical explanation remains that of $\eta$ and $N$ exchange 
as proposed in Ref.[6].

Radiative lepton-flavor-changing decays (i.e. $\tau \to \mu \gamma$, 
$\tau \to e \gamma$, $\mu \to e \gamma$) through $\eta$ and $N$ 
exchange are suppressed by the near degeneracy of the neutrino mass matrix, 
as explained in Ref.[6].  However, they also receive contributions from 
Eq.~(31).  The most prominent process is actually $\mu \to e \gamma$ 
from $(\phi')^0$ exchange, with an amplitude given by
\begin{equation}
{\cal A} = {1 \over 32 \pi^2} {m_\tau^2 \over M_1^2} {m_\mu \over v^2} 
\epsilon^\alpha q^\beta \bar e \sigma_{\alpha \beta} \left( {1+\gamma_5 
\over 2} \right) \mu.
\end{equation}
Hence
\begin{equation}
B(\mu \to e \gamma) = {9 \over 32 \pi^2} {m_\tau^4 \over M_1^4} \left( 
{v_0^2 \over 3 v^2} \right)^2.
\end{equation}
Using the experimental upper bound \cite{meg} of $1.2 \times 10^{-11}$, 
we find
\begin{equation}
M_1 = {M_R M_I \over |M_R^2 - M_I^2|^{1/2}} > 390~{\rm GeV} \left( {v_0 
\over \sqrt 3 v} \right).
\end{equation}

\section{Quark Sector}

In the quark sector, we could also try having the three left-handed quark 
doublets transform as \underline {3} under $A_4$, and the right-handed quark 
singlets as \underline {1}, \underline {1}$'$, and \underline {1}$''$.  In 
that case, the quark mass matrices corresponding to Eq.~(19) are diagonal just 
as that of the charged leptons.  Since the soft breaking of $A_4$ is not 
possible in the quark sector, the only way that a charged-current mixing 
matrix may arise is from the violation of $v_1 = v_2 = v_3$.  However, 
because the mixing is further suppressed by the ratio of quark masses, 
the final effect is negligible.

Suppose we assign both quark doublets and singlets to be \underline {3} 
under $A_4$.  Then there are 2 invariant couplings to $\Phi_i$ as shown 
by Eq.~(1).  However, the mass eigenvalues in this case are those of 
Eq.~(27), which do not match the observed quark masses.

To accommodate realistic quark mass matrices with the correct charged-current 
mixing matrix, we can just go back to the standard model, i.e. all quarks 
are trivial under $A_4$ as well as another Higgs doublet $\Phi_4$.  Thus 
\begin{equation}
v_0^2 = v_1^2 + v_2^2 + v_3^2 + v_4^2 + u^2 = 3 v^2 + v_4^2 + u^2.
\end{equation}

\section{Concluding Remarks}

In conclusion, we have shown how nearly degenerate neutrino masses can be 
obtained in the context of a softly and spontaneously broken discrete 
$A_4$ (tetrahedral) symmetry while allowing realistic charged-lepton and 
quark masses.  In addition to the standard-model particles, we have 3 heavy 
neutral right-handed singlet fermions $N_i$ at the TeV scale or below, 
whose decay into charged leptons would map out the neutrino mass matrix 
as discussed in Ref.[5].  The nearly mass-degenerate $N_i$ can explain 
the possible discrepancy of the muon anomalous magnetic moment as discussed 
in Ref.[6].  The 3 new Higgs scalar doublets $\Phi_i$ of this model have 
distinct experimental signatures.  One combination, i.e. $(\Phi_1 + \Phi_2 + 
\Phi_3)/\sqrt 3$ behaves like the standard-model Higgs doublet, except that 
it couples only to leptons.  The other two, i.e. $\Phi'$ and $\Phi''$ of 
Eq.~(32), are predicted to be pairwise mass-degenerate and have precisely 
determined flavor-changing couplings as given by Eq.~(31).  They are 
consistent with all present experimental bounds and amenable to 
experimental discovery below a TeV.

\section*{Acknowledgements}

This work was supported in part by the U.~S.~Department of Energy
under Grant No.~DE-FG03-94ER40837.  G.R. also thanks the UCR Physics 
Department for hospitality.

\bibliographystyle{unsrt}

\begin{thebibliography}{99}
\bibitem{tmg} As shown in Ref.\cite{mr}, a hierarchical neutrino mass matrix 
would then confict with the experimental upper bound on the $\tau \to \mu 
\gamma$ rate.
\end{thebibliography}

\end{document}